\documentclass[preprint]{aastex}
\shorttitle{Scalable Correlator Architecture}
\shortauthors{Parsons et al.}

\usepackage{amsmath}
\usepackage{graphicx}
\usepackage{natbib}
\begin{document}
\title{A Scalable Correlator Architecture Based on 
    Modular FPGA Hardware, Reuseable Gateware, and Data Packetization}

\author{Aaron Parsons, Donald Backer, and Andrew Siemion}
\affil{Astronomy Department, 
    University of California, Berkeley, CA}
\email{aparsons@astron.berkeley.edu}
\author{Henry Chen and Dan Werthimer}
\affil{Space Science Laboratory,
    University of California, Berkeley, CA}
\author{Pierre Droz, Terry Filiba, Jason Manley\altaffilmark{1}, 
    Peter McMahon\altaffilmark{1}, and Arash Parsa}
\affil{Berkeley Wireless Research Center,
    University of California, Berkeley, CA}
\author{David MacMahon, Melvyn Wright}
\affil{Radio Astronomy Laboratory,
    University of California, Berkeley, CA}

\altaffiltext{1}{Affiliated with Karoo Array Telescope,
    Cape Town, South Africa}

\begin{abstract}
A new generation of radio telescopes is achieving unprecedented levels of
sensitivity and resolution, as well as increased agility and field-of-view, by
employing high-performance digital signal processing hardware to phase and
correlate large numbers of antennas.  The computational demands of these
imaging systems scale in proportion to $BMN^2$, where $B$ is the signal
bandwidth, $M$ is the number of independent beams, and $N$ is the number of
antennas.  The specifications of many new arrays lead to demands in excess of
tens of PetaOps per second.

To meet this challenge, we have developed a general purpose correlator
architecture using standard 10-Gbit Ethernet switches to pass data
between flexible hardware modules containing Field Programmable Gate Array
(FPGA) chips.  These chips are programmed using open-source signal processing
libraries we have developed to be flexible, scalable, and chip-independent.
This work reduces the time and cost of implementing a wide range of signal
processing systems, with correlators foremost among them, and facilitates
upgrading to new generations of processing technology. We present several
correlator deployments, including a 16-antenna, 200-MHz bandwidth, 4-bit, full
Stokes parameter application deployed on the Precision Array for Probing the
Epoch of Reionization.
\end{abstract}


\section{Introduction}
\label{sec:intro}

Radio interferometers, which operate by correlating the signals from two or
more antennas, have many advantages over traditional single-dish telescopes,
including greater scalability, independent control of aperture size and
collecting area, and self-calibration.  Since the first digital correlator
built by Weinreb \citep{weinreb_1961}, the processing power of
these systems has been tracking the Moore's Law growth of digital electronics.
The decreasing cost per performance of these systems has influenced the design
of many new radio antenna array telescopes.  Some
next-generation array telescopes at meter, centimeter and millimeter
wavelengths are: 
the LOw Frequency ARray (LOFAR), 
the Precision Array for Probing the Epoch of Reionization (PAPER), 
the Murchison Widefield Array (MWA), 
the Long Wavelength Array (LWA),
the Expanded Very Large Array (EVLA), 
the Allen Telescope Array (ATA), 
the Karoo Array Telescope (MeerKAT), 
the Australian Square Kilometer Array Demonstrator (ASKAP),
the Atacama Large Millimeter Array (ALMA).
and the Combined Array for Research Millimeter-wave Astronomy (CARMA). 
This paper presents a novel approach to the intense digital signal 
processing requirements of these instruments that has many other applications
to astronomy signal processing.

While each generation of electronics has brought new commodity data processing
solutions, the need for high-bandwidth communication between processing nodes
has historically lead to specialized system designs.  This communication
problem is particularly germane for correlators, where the number of
connections between nodes scales with the square of the number of antennas.
Solutions to date have typically consisted of specialized processing boards
communicating over custom backplanes using non-standard protocols.  However,
such solutions have the disadvantage that each new generation of digital
electronics requires expensive and time-consuming investments of engineering
time to re-solve the same connectivity problem.  Redesign is driven by the same
Moore's Law that makes digital interferometry attractive, and is not unique to
the interconnect problem; processors such as Application-Specific Integrated
Circuits (ASICs) and Field Programmable Gate Arrays (FPGAs) also require
redesign, as do the boards bearing them, and the signal processing algorithms
targeting their architectures.

Our research is aimed at reducing the time and cost of correlator design and
implementation.  We do this, firstly, by developing a packetized communication
architecture relying on industry-standard Ethernet switches and protocols to
avoid redesigning backplanes, connectors, and communication protocols.
Secondly, we develop flexible processing modules that allow identical boards to
be used for a multitude of different processing tasks.  These boards are
applicable to general signal processing problems that go beyond 
correlators and even radio science to include, e.g., ASIC design and
simulation, genomics, and research into parallel processor architectures.  
General
purpose hardware reduces the number of boards that have to be redesigned and
tested with each new generation of electronics.  Thirdly, we create
parametrized signal processing libraries that can easily be recompiled and
scaled for each generation of processor.  This allows signal processing systems
to quickly take advantage of the capabilities of new hardware.  Finally, we
employ an extension of a Linux kernel to interface between CPUs and FPGAs for
the purposes of testing and control, presenting a standard file interface
for interacting with FPGA hardware.

This paper begins with a presentation of the new correlator
design architecture in \S\ref{sec:architecture}. The hardware to 
implement this architecture follows in \S\ref{sec:hardware}, and
the FPGA gateware used in the hardware is summarized in \S\ref{sec:gateware}.
Issues concerning system integration are given in \S\ref{sec:integration},
and performance characterization of subsystems are given in 
\S\ref{sec:characterization}. Results from our first deployments of
the packetized correlator are displayed in \S\ref{sec:deployments}.
Our final section summarizes our progress and points to a number
of directions we are pursuing for the next generation of scalable
correlators based on modular hardware, reuseable gateware and
data packetization. An appendix gives a glossary of technical
acronyms since this paper makes heavy use of abbreviated terms.

\section{A Scalable, Asynchronous, Packetized FX Correlator Architecture}
\label{sec:architecture}

Correlators integrate the pairwise correlation between complex voltage samples 
from polarization channels of array antenna receivers at a set of
frequencies.
Once instrumental effects have been calibrated and removed, the resultant 
correlations (called visibilities) represent the self-convolved electric field
across an aperture sampled at locations 
corresponding to separations between antennas.  These visibilities can be 
used to reconstruct an image of the sky by inverting the interferometric 
measurement equation:
\begin{equation}
V_{\nu}(u,v)=\int\!\!\!\!\int{G_{i,\nu}G_{j,\nu}^*I_\nu(\ell,m)}
{e^{-2\pi i(u\ell+vm+w(\sqrt{1-\ell^2-m^2}-1))}d\ell dm}
\label{eq:vis}
\end{equation}
$I_\nu$ represents the sky brightness in angular coordinates $(\ell,m)$, and 
$(u,v,w)$ correspond to the separation in wavelengths of an antenna pair 
relative to a pointing direction.
For antennas with separate polarization feeds, cross-correlation
of polarizations yields components of the four Stokes parameters that
characterize polarized radiation, here defined in terms of linear
polarizations ($\|,\perp$) for all pairs of antennas $A$ and $B$ 
\citep{rybicki_lightman1979}:
\begin{equation}
\begin{array}{ll}
\displaystyle I=A_\| B_\|^*+A_\perp B_\perp^* &\ \ \ 
Q=A_\| B_\|^*-A_\perp B_\perp^* \nonumber \\
\displaystyle U=A_\| B_\perp^*+A_\perp B_\|^* &\ \ \ 
V=A_\| B_\perp^*-A_\perp B_\|^* 
\label{eq:pol}
\end{array}
\end{equation}
I measures total intensity, V measures the degree of circular polarization,
and Q and U measure the amplitude and orientation of linear polarization.

The problem of computing pairwise correlation as a function of frequency can be
decomposed two mathematically equivalent but architecturally distinct ways.
The first architecture is known as ``XF'' correlation because it first
cross-correlates antennas (the ``X'' operation) using a time-domain ``lag''
convolution, and then computes the spectrum (the ``F'' operation) for each resulting
baseline using a Discrete Fourier Transform (DFT).  An alternate architecture
takes advantage of the fact that convolution is equivalent to multiplication in
Fourier domain.  This second architecture, called ``FX'' correlation, first
computes the spectrum for each individual antenna (the F operation), and then
multiplies pairwise all antennas for each spectral channel (the X operation).  
An FX correlator has an advantage over XF
correlators in that the operation that scales as $O(N^2)$ with the
number of antennas, N, is a complex multiplication as opposed to a full 
convolution in an XF correlator \citep{daddario2001,yen1974}.

Though there are mitigating factors (such as bit-growth for representing the
higher dynamic range of frequency-domain data) that favor XF correlators for
small numbers of antennas \citep{thompson_et_al2001}, FX correlators are more
efficient for larger arrays.  Since scalability to large numbers of antennas is
one of the primary motivations of our correlator architecture, we have chosen
to develop FX architectures exclusively.

\subsection{Scalability With Number of Antennas and Bandwidth}
\label{sec:scalability}

The challenge of creating a scalable FX correlator is in designing a
scalable architecture for factoring the total computation into manageable 
pieces and efficiently bringing together data in each piece for computation.
Traditionally, the spectral decomposition (in F engines) has been scaled to
arbitrary bandwidths by using analog mixers and filters to divide the operating
band of each antenna into the widest subbands that can be processed digitally
using existing technology.  Within correlation of a given subband,
the complexities of computation and of data distribution both scale
linearly with bandwidth and quadratically with the number of antennas. It is
imperative that the arrangement of cross-multiplication engines (hereafter
referred to as X engines) minimize data replication/retransmission, even as X
engines expand to encompass many boards.  Fortunately, each frequency channel
of an FX correlator is computationally independent, providing a natural
boundary for dividing computation among processing nodes.

\begin{figure}
\begin{center}
\includegraphics[scale=.4]{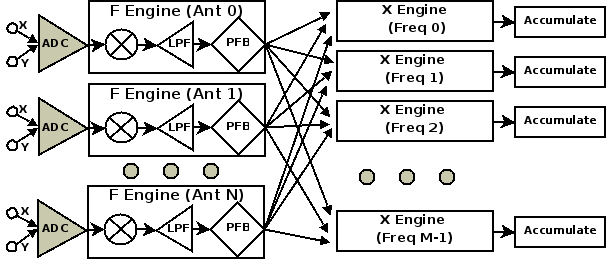}
\caption{In a simplistic FX correlator,
the signals from N antennas are first decomposed into M frequency channels 
(F operation) and then cross-multiplied (X operation).  Different channels are
never cross-multiplied, making them natural units for X engine processing.
Thus, each X engine handles all baselines for one frequency channel.
\label{fig:corr_arch1}}
\end{center}
\end{figure}

\placefigure{fig:corr_arch1}

Figure \ref{fig:corr_arch1} illustrates a simplistic architecture for an FX
correlator that takes advantage of the computational independence of channels
to avoid unnecessary data transmission;
the total X computation has been factored into X engines that cross-multiply 
all antenna pairs for a single frequency channel.
This architecture is overly
simplistic, since an X engine's performance can be equated to an aggregate 
input bandwidth that it can handle.  For the sake of efficiency, an X engine 
processor 
should receive as many channels as it has capacity to process.  In this case, 
the number of X engines is given by:
\begin{equation}
\#\ {\rm X\ Engines} = \frac{({\rm Antenna\ Bandwidth})\times 
  (\#\ {\rm Antennas})}{ {\rm X\  Engine\ Processing\ Bandwidth}}
\end{equation}
Multiplexing channels into X engines makes cross-multiplication
complexity independent of the number of channels.  There are three
potential bottlenecks for scaling this architecture: the complexity of
interconnecting F engines and X engines, the bandwidth into individual X
engines, and the amount of computation in an X engine relative to the size of a
processing chip/board/system.  Each of these bottlenecks warrants further
discussion.

The potential bottleneck of connecting $N$ antenna-based F engines to $M$
channel-based X engines is highlighted by the criss-crossed lines in Figure
\ref{fig:corr_arch1}.  Historically, this bottleneck has been addressed with
custom backplanes and transmission protocols.  However, our group has taken the
novel approach of using high-performance, commercially available, 
10-Gbit/s Ethernet (10GbE) switches to solve this problem.  
As will be discussed, these switches currently have the bandwidth and switching
capacity to handle large correlators, and represent a negligible fraction of
the total cost of correlator hardware.  Furthermore, switching technology is
driven by commercial applications and by Moore's Law, making it likely that
future switches will continue increasing in number of ports and bandwidth per
port.

A second potential bottleneck concerns how data rates and 
numbers of X engines scale with antenna bandwidth.  It is important that
we consider various bandwidth cases, owing to the variety of science
applications driving large, next-generation systems.  For example, correlators
for large arrays of low-bandwidth antennas will need to multiplex data into
higher bandwidth processors, while arrays with larger bandwidths will face the
opposite problem. In our architecture, we make the reasonable assumption that
the number of frequency channels always exceeds the number of antennas.  
This assumption
ensures that the per-port bandwidth into an X engine never exceeds what is
transmitted per antenna.  Multiple channels may then be mapped into an X engine
up to its computational capacity (allowing efficient resource utilization for
low-bandwidth arrays), and additional X engines may be added for high-bandwidth
applications.  Antenna bandwidths requiring transmission above 10 Gbits/s can
be accommodated by connecting F engines to multiple 10GbE ports.
Frequency channels are then assigned to each port, which connect separate
switches and sub-networks of X engines.  In this way, bandwidths may be scaled
up to the transmission capability of an F processor by increasing the number of
subnets, and not switch complexity.

The third and final potential bottleneck concerns how the sizes of individual X
engines scales with the number of antennas.  Both large and small numbers of
antennas pose scaling problems.  The size of an X engine responsible for
computing all baseline cross-multiples with a fixed input data rate
scales as $O(N)$, while
the number of X engines required to accommodate the expanding data bandwidth
with increasing numbers of antennas also scales as $O(N)$,
accounting for the $O(N^2)$ scaling of computing in a correlator.  For
sufficiently large $N$, the size of an X engine can exceed the size of any
processing chip or board.  Our solution has been to develop an X engine whose
pipelined architecture allows it to be split across multiple processors with
simple point-to-point connectivity.  This allows many processors to be chained
together from a switch port to meet the computational demands of an X engine.
Scaling to small $N$ is equally challenging, because the aggregate correlator
bandwidth decreases as $O(N)$, while computational complexity scales down as
$O(N^2)$.  As a result, we can find that the number of X engines that
fit onto a chip/board exceeds the rate at which data can be received.  The
threshold where this problem is encountered can be changed by designing
processors with greater connectivity, but once hardware is fixed, there is no
other recourse but to accept a certain inefficiency for low numbers of
antennas.  While this is a fundamental limitation of our architecture, 
the cost of small correlators is typically dominated by development
(not hardware), so a certain architectural inefficiency can be accommodated for
the savings it affords in development time.

\subsection{Globally Asynchronous Locally Synchronous Systems}
\label{sec:gals}

Packetized data transmission simplifies the cross-connect problem inherent to
correlators, but this comes at the price of global synchronicity.  Packetized
communication is fundamentally asynchronous: data can arrive scrambled,
delayed, or not at all.  Locally-synchronous X engine processing must therefore
transition from being timing-driven (with throughput tied to an FPGA clock, for
example) to being asynchronously data-driven.  Though data buffers and control
signals complicate development, Globally Asynchronous Locally Synchronous
(GALS) design facilitates system integration and leads to robust design
\citep{chapiro1984,luis_et_al2007}.  Processors run at clock rates above the
data rate, using local oscillators that can drift with temperature.  By
allowing for non-transmission of data, individual components can fail without
causing global failure--an important feature for large systems where
components may fail regularly during operation.  GALS design also insulates
processing architectures from decisions regarding sample rates and antenna
bandwidths, allowing for greater operational flexibility.  Finally, individual
processing elements may be redesigned and upgraded in a GALS system without
affecting the overall architecture, facilitating early adoption of new
technology.

Data-driven processing on locally synchronous processors like FPGAs requires
controlling propagation through the processing pipeline.  However, routing
control signals to every multiplier, accumulator, and logic element in a
pipeline can lead to excessive routing and gating demands.  To avoid this, we
have implemented a window-based processing architecture for algorithms where
the results derived from one set of data samples are computationally
independent from the next.  In this architecture, processing elements are
allowed to run freely at their native rate without being enabled/disabled, but
are only provided data when an entire window of data has been buffered.  These
windows of data are provided synchronously with the inherent window boundaries
of the processing element, and an entire output window is flagged as valid.
Internally, a processor processes both valid and invalid data--it is only the
external buffering system that keeps track of data validity.  This technique is
applicable to many common operators such as cross-multipliers, DFTs, and
accumulators.  Finite Impulse Response (FIR) filtering is an
operation notable for not being window-based.

\subsection{Example Applications}
\label{sec:example}

\begin{figure}
\begin{center}
\includegraphics[scale=.25]{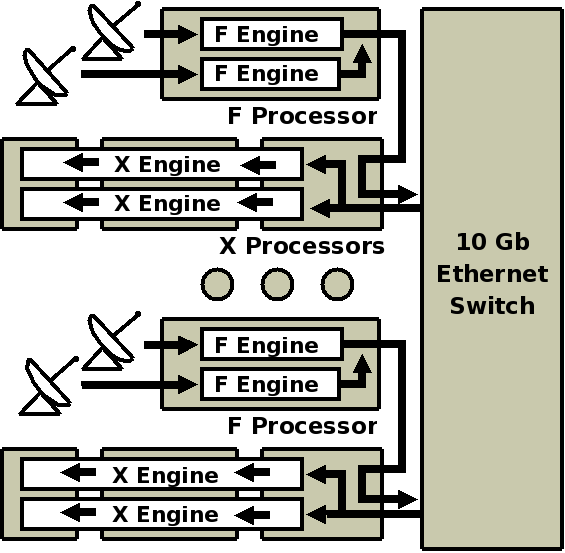}
\caption{Data bandwidth per antenna is equal to the processing bandwidth of 
an X processor in this example application.  Transmitted data is routed 
through an X processor to take advantage of bidirectionality of 10GbE ports, 
thereby halving the number of ports on the switch.
\label{fig:ex_app1}}
\end{center}
\end{figure}

\placefigure{fig:ex_app1}

Perhaps the best method for demonstrating the flexibility and scalability of
our correlator architecture is through example applications.  To illustrate
techniques for using hardware and ports efficiently, we will map processing
into fictitious hardware that corresponds roughly in capability to the
CASPER (Center for Astronomy Signal Processing 
and Engineering Research)\footnote{http://casper.berkeley.edu}
hardware discussed in Section \ref{sec:hardware}.

Our first example (Fig. \ref{fig:ex_app1}) illustrates an antenna signal
bandwidth sufficiently low so that data from 2 polarization channels of 2
antennas can be transmitted over one 10GbE connection.  Assuming that the
number of antennas evenly divides the number of frequency channels, and that
the processing bandwidth of an X engine matches the data bandwidth of one
antenna, there will be the same number of X engines as F engines, and each X
engine will receive 1/N$^{\rm th}$ of the total bandwidth, where N is
the number of antennas.  F engine
transmission and X engine reception are combined on a single port to make use
of the bi-directionality of 10GbE.  This optimization halves the size of the
switch needed.  Multiple X processors can be chained together from a single
10GbE port using point-to-point connections.  For cases where the number of
antennas does not evenly divide the number of frequency channels, one can adjust
packet transmission to drop remainder channels so that the band may be equally
divided among X engines.

\begin{figure}
\begin{center}
\includegraphics[scale=.25]{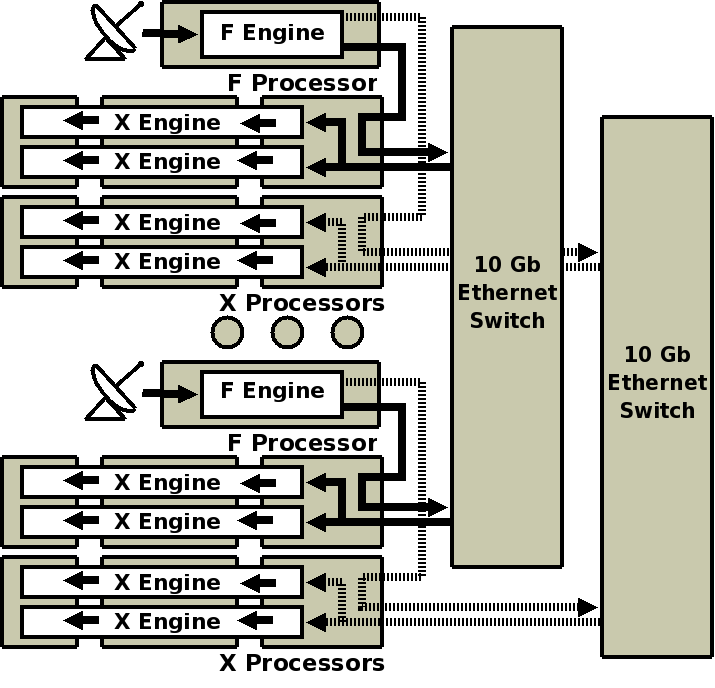}
\caption{Data bandwidth per antenna can exceed
what can be carried over 10GbE.  Here, the frequency band has been spread 
across ports by channel, so that each half of transmission occurs on an 
isolated subnet.  This is possible because different channels are never 
cross-multiplied in an FX correlator.
\label{fig:ex_app2}}
\end{center}
\end{figure}

\placefigure{fig:ex_app2}

A second example (Fig. \ref{fig:ex_app2}) illustrates a case where the
bandwidth from a single F engine exceeds the transmission capacity of a 10GbE
link.  Here, data can be split by frequency channel across two
ports.  Since different channels are never cross-multiplied, each of these
links goes to a separate subnet of switched X engines.  Thus,
two smaller (and often less expensive per port) switches may be 
substituted for one large
one.  Each X engine still receives the same bandwidth as in the previous
example, although this now represents a smaller fraction of the total
bandwidth.  Note that the same X processor used in the first example functions
here without modification.  Only the number of X engines and the transmission
pattern has changed.

\begin{figure}
\begin{center}
\includegraphics[scale=.25]{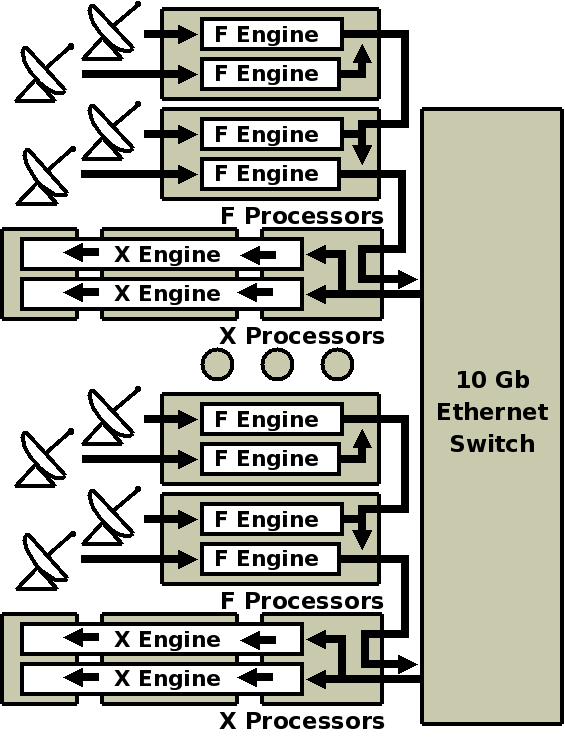}
\caption{When the processing bandwidth of an X engine exceeds the antenna 
bandwidth by at least a factor of 2, half as many X processors are needed for 
a given number of antennas.  X processors operate independently of data
bandwidth; the same design handles this and the previous two cases 
(Figs. \ref{fig:ex_app1} and \ref{fig:ex_app2}).  Only the number of X 
processors and the data transmission pattern have changed.
\label{fig:ex_app3}}
\end{center}
\end{figure}

\placefigure{fig:ex_app3}

A final example (Fig. \ref{fig:ex_app3}) explores the case where the capacity
of an X processor and a 10GbE link both exceed the data bandwidth.  In this
case, multiple F engines can (but do not have to) be chained together to
minimize the number of switched ports.  As should be the case, only half as
many X engines (as compared to Fig. \ref{fig:ex_app1}) are necessary for a
given number of antennas.  X processors operate in the same configuration as
before, oblivious to changes in F engines.

These examples highlight the flexibility of the hardware and gateware for
targeting a number of applications. One shortcoming they also illustrate is
how the cabling between components differs for different bandwidths.
Therefore the different bandwidth operations are not as easily reconfigured as
might be desired for varying science goals on a given telescope. Research is
ongoing to improve the rapid reconfigurability that is an essential
specification for the most general radio interferometer array applications.

\section{Modular, FPGA-based Processing Hardware}
\label{sec:hardware}

A flexible and scalable correlator architecture is of limited use without
equally dynamic processing hardware that can support a variety of
configurations.  FPGAs provide a unique combination of flexibility and
performance that make them well-suited for moderate-scale signal processing
applications such as correlators and spectrometers \citep{parsons_et_al2006}.
A primary goal of the CASPER group has been development of
multipurpose processing modules that can be of general use to the astronomy
signal processing community, and beyond.  We seek to
minimize the effort of redesigning and upgrading hardware by modularizing
processing hardware, by minimizing the number of different modules 
in a system, and by employing industry-standard interconnection protocols.

Hardware modularity is the idea that boards should have consistent interfaces
in order to be connectible with an arbitrary number of heterogeneous components
to meet the computing needs of an application (``computing by the yard''), and
that upgrading/revising a component does not change the way in which components
are combined in the system.
Minimization of hardware reproduction costs is often used to motivate the
design of specialized hardware for large-scale correlators.  However, 
the longer development times inherent to such solutions, and
the necessity of targeting specific components from the outset,
suggest that a modular solution, initiated nearer to the deployment date,
will employ newer technology that costs less and uses less
power per operation.  The predicted economy of mass-producing
specially-designed hardware must be tempered by its expected devaluation
by Moore's Law over the course of correlator development.  This devaluation
makes the argument that hardware modularity can reduce the overall system
cost, even for large-scale systems, by reducing development time.

In current correlator systems, we rely on two
CASPER FPGA-based processing boards; Internet Break-Out Boards (IBOBs) are
generally used for implementing per-antenna F engine processing, and
second-generation Berkeley Emulation Engines (BEE2s) implement X engine
processing.  Work is progressing on a new board, the Reconfigurable Open
Architecture for Computing Hardware (ROACH), that will provide a single-board
solution to both F and X processing.

\begin{figure}
\begin{center}
\includegraphics[scale=.25]{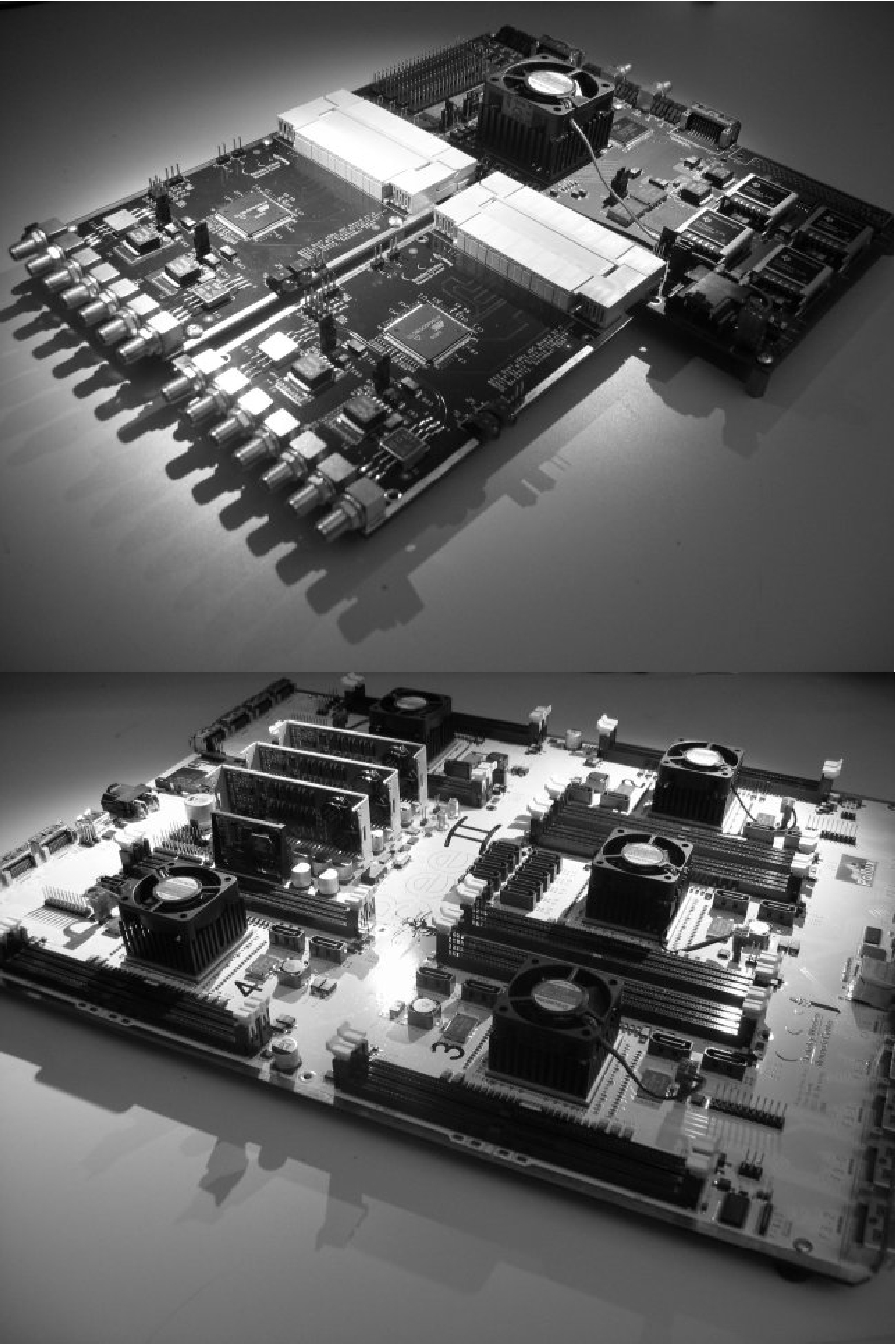}
\caption{%
Our correlator architecture relies on modular FPGA-based processing hardware 
developed by our group to
combine flexibility, upgradeability, and performance.  Illustrated above are:
(top) IBOB and ADC FPGA/digitizer modules
(bottom) The Berkeley Emulation Engine (BEE2) FPGA board
\label{fig:ibobadcbee2}}
\end{center}
\end{figure}

\placefigure{fig:ibobadcbee2}

IBOBs (Fig. \ref{fig:ibobadcbee2}) can interface to two
Analog-to-Digital Converter (ADC) boards, each capable of digitizing two
streams at 1 Gsamples/sec or a single stream at 2 Gsamples/sec using an Atmel
AT84AD001B dual 8-bit ADC chip.  This data is processed by a Xilinx XC2VP50
FPGA containing 232 18$\times$18-bit multipliers, two PowerPC CPU cores, and
over 53,000 logic cells.  Two ZBT SRAM chips provide 36 Mbits of extra
buffering, and two 10GbE-compatible CX4 connectors provide a standard interface
for connecting to other boards, switches, and computers.  A detailed discussion
of ADC signal fidelity is presented in Section \ref{sec:characterization}.
We are developing a second ADC board that allows four signal sampling at
200 Msample/sec. 

The BEE2 board \cite{chang_et_al2005} (Fig. \ref{fig:ibobadcbee2}) was
originally designed for high-end reconfigurable computing applications such as
ASIC design, but has been conscripted for astronomy applications in a
collaboration between the BWRC\footnote{Berkeley Wireless Research Center
http://bwrc.eecs.berkeley.edu}, 
the UC Berkeley Radio Astronomy Laboratory, and the UC Berkeley SETI group. 
The 500
Gops/sec of computational power in the BEE2
is provided by 5 Xilinx XC2VP70 Virtex-II Pro
FPGAs, each containing 328 multipliers, two PowerPC CPU cores capable of
running Linux, and over 74,000 configurable logic cells.  Each FPGA connects to
4 GB of DDR2-SDRAM, and four 10GbE-compatible CX4 connectors, and all FPGAs
share a 100-Mbps Ethernet port.  The size and connectivity of the
BEE2 board make it suitable for implementing X engine processing in our
correlator architecture.

The ROACH board is being developed in collaboration with MeerKAT and 
NRAO,\footnote{The National Radio Astronomy
Observatory (NRAO) is owned and operated by Associated Universities, Inc. with
funding from the National Science Foundation} 
and is scheduled for release in the third quarter of 2008.  It is intended as a
replacement for both IBOB and BEE2 boards.  A single Xilinx Virtex-5 XC5VSX95T
FPGA containing 94,000 logic cells and 640 multiplier/accumulators provides 400
Gops/sec of processing power and is connected to a separate PowerPC 440EPx
processor with a 1 GbE network connection.  The board contains 4 GB of DDR2
DRAM and two 36Mbit QDR SRAMs, four 10GbE-compatible CX4 connectors, and two
interfaces that allow the use of the current ADC boards, or a new 3
Gsamples/sec (6 Gsamples/sec dual-board interleaved) ADC.  The scale, economy,
and peripheral interfaces of this board will make it appropriate for both F and
X engine processing, and will enable a single-board correlator architecture.

\placetable{tab:hardware_price}

\section{Gateware}
\label{sec:gateware}

Efficient, customizable signal processing libraries are another important
component of a flexible and scalable correlator architecture.  Towards this
goal, our group has designed a set of open-source libraries\footnote{Available
at http://casper.berkeley.edu} for the Simulink/Xilinx System Generator FPGA
programming language.  These libraries abstract chip-specific components to
provide high-level interfaces targeting a wide variety of devices.  Signal
processing blocks in these libraries are parametrized to scale up and down to
arbitrary sizes, and to have selectable bit widths, latencies, and scaling.
Though the design principles of parametrization and scalability have added
complexity to the initial design of these libraries, it dramatically enhances
their applicability and potential for longevity as hardware evolves.  It also
decreases testing time by allowing developers to debug scale models of systems
that derive from the same parametrization code and are behaviorally similar to
larger systems.  In this section, we present several components of our
libraries vital to the design of flexible correlators.

\subsection{A Digital Down-Converter}
\label{sec:downconverter}

The rising speed of ADCs has enabled digitization to occur increasingly early
in the antenna receiver chain.  We are thus replacing analog electronics
commonly known as intermediate frequency processor (gain, band definition)
and baseband mixer (conversion to zero frequency and filtering).
There are numerous advantages to doing this.
Digital mixing allows dynamically selecting an operating frequency within the
digitized band while ensuring perfect sine-cosine phasing in the local
oscillator (LO) mixing frequency. 
Digitizing a wider bandwidth than will be ultimately processed makes analog
filtering less critical; inexpensive filters with slow roll-offs can be
used, and passband rippling can be corrected.  Finally, digital filtering
allows flexibility and control in selecting passband shapes and adjusting fine
delays.  One can even split out several bands from the same signal.
The issue of quantization levels and other digital artifacts needs to be
carefully addressed.

Our library provides a digital down-conversion core with a runtime-selectable
mixing frequency.  Using a discretely sampled sine wave in an addressable
lookup table, we can approximate nearly any mixing frequency by rounding a wide
accumulation register (incremented every clock) to the nearest address in the
lookup table.  Digital sine waves have an accuracy dictated by the number of
bits used to represent a value; a lookup table need only have enough samples to
achieve comparable accuracy.  The fact that the derivative of $\sin(x)$ reaches 
a maximum magnitude of 1 allows the sampling interval of a sine wave to be
simply equated to the accuracy of a coefficient over that time interval.
As a result, a lookup table only need be addressed with the same
bit-width as the sample width to implement an arbitrary mixing frequency.

\begin{figure}
\begin{center}
\includegraphics[scale=.45]{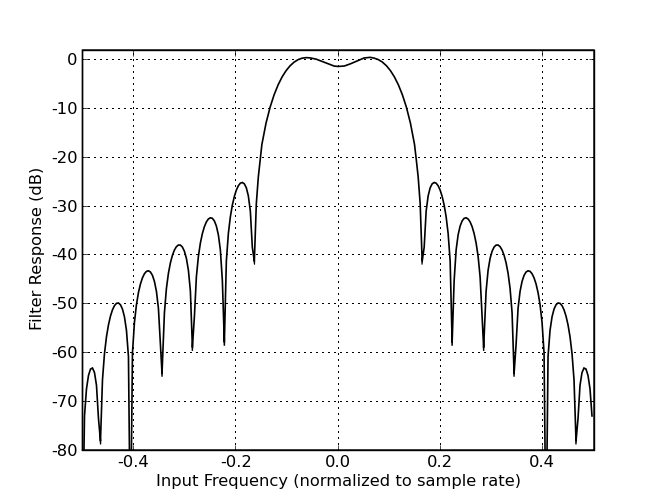}
\caption{%
This example response an the FIR filter in a digital down-converter, 
illustrates the 16 tap low-pass design used in the correlator deployments
presented later.
\label{fig:ddc_passband}}
\end{center}
\end{figure}

\placefigure{fig:ddc_passband}

Our library also contains a decimating FIR filter.  Digital filters have
advantages over analog filters by being reprogrammable and by providing exact,
calculable passbands.  This filter is often used for suppressing harmonics of
the mixing frequency and for steepening the rolloff of cheaper analog filters,
but it has also been relied upon for implementing IF sub-band selection
digitally.  In practice, one must weigh the need for performance and
flexibility against the cost of FPGA resources compared to analog filters.  As
an example, the response of the FIR filter used in various correlator designs
is shown in Figure \ref{fig:ddc_passband}.  Since the exact shape 
of this filter can be calculated, it is possible to remove passband
ripple post-channelization because of the large dynamic range available in
output of our FFT core.

\subsection{A Polyphase Filter Bank Front-End}
\label{sec:pfb}

\begin{figure}
\begin{center}
\includegraphics[scale=.52]{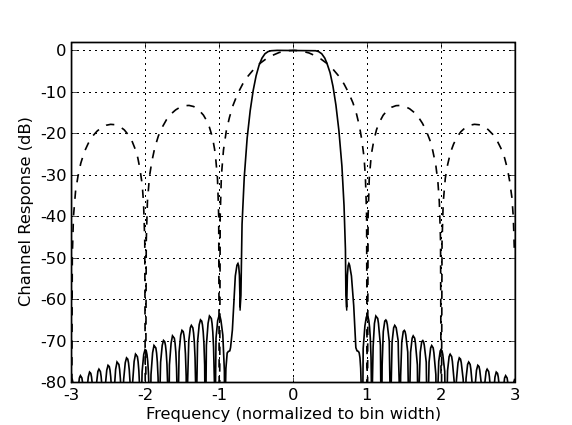}
\caption{%
The response of a frequency channel in an 8-tap Polyphase Filter Bank (solid)
using a Hamming window is compared to an equivalently sized Discrete Fourier 
Transform (dashed).  This particular PFB, implemented for 2048 channels, is 
used in the correlator deployments presented in Section \ref{sec:deployments}.
\label{fig:pfb_bin_resp}}
\end{center}
\end{figure}

The Polyphase Filter Bank (PFB) \citep{crochiere+rabiner1983, vaidyanathan1990}
is an efficient implementation of a bank of evenly spaced, decimating FIR
filters.  The PFB algorithm decomposes these filters into a single polyphase
convolution followed by a DFT.  Since DFTs have been highly optimized
algorithmically, this results in an extremely efficient implementation.
Equivalently, the PFB may be regarded as an improvement on the Fast Fourier
Transform (FFT) that uses a front-end polyphase FIR filter to improve the
frequency response of each spectral channel (Fig. \ref{fig:pfb_bin_resp}).
This improvement comes at the cost of buffering an additional window of samples
and adding a complex cross-multiplication for each additional tap in the
polyphase FIR.  This PFB implementation has seen widespread use in the astronomy
community in 21 cm hydrogen surveys \citep{heiles_et_al2004}, pulsar surveys
\citep{demorest_et_al2004}, antenna arrays \citep{bradley_et_al2005}, Very Long
Baseline Interferometry, and other applications.

\placefigure{fig:pfb_bin_resp}

Our core is parametrized to use selectable windowing functions, allowing
adjustment of the out-of-band rejection and passband ripple/rolloff.  Blackman
and Tukey \citep{blackman_tukey1958} provides a summary of the characteristics
and trade-offs of various windows.  Each polyphase FIR tap, at the cost of
increased buffering and additional multipliers, increases filter steepness by
adding samples (in increments of the number of channels) to the time window
used in the PFB.  For fixed-point implementations, a practical upper limit to
the number of PFB taps is set by the number of bits used to represent filter
coefficients; the sinc function's 1/x tapering ceases to be representable when
$\pi T > \pi + 2^{B+1}$ where $T$ is the number of taps, and $B$ is the
coefficient bit width.  Finally, the width of a PFB channel is tunable by
adjusting the period of the sinc function, forcing adjacent bandpass filters to
overlap at a point other than the -3 dB point.  Note that this causes
power to no longer be conserved in the Fourier transform operation.

\subsection{A Bandwidth-Agile Fast Fourier Transform}
\label{sec:fft}

The computational core of our FFT library is an implementation of a radix-2
biplex pipelined FFT \citep{rabiner_gold1975} capable of analyzing two
independent, complex data streams using a fraction of the FPGA resources of
commercial designs \citep{dick2000}.  This architecture takes advantage of the
streaming nature of ADC samples by multiplexing the butterfly computations of
each FFT stage into a single physical butterfly core.  When used to analyze two
independent streams, every butterfly in this biplex core outputs valid data
every clock for 100\% utilization efficiency.

The need to analyze bandwidths higher than the native clock rate of an FPGA led
us to create a second core that combines multiple biplex cores with additional
butterfly cores to create an FFT that is parametrized to handle $2^P$ samples
in parallel \citep{parsons2008}.  This FFT architecture uses only 25\% more
buffering than the theoretical minimum, and still achieves 100\% butterfly
utilization efficiency.  This feat is achieved by decomposing a $2^N$
channel FFT into $2^P$ parallel biplex FFTs of length $2^{N-P}$, followed by a
$2^P$ channel parallel FFT core using time-multiplexed twiddle-factor
coefficients.

Finally, we have written modules for performing two real FFTs with each half of
a biplex FFT using Hermitian conjugation.  Mirroring and
conjugating the output spectra to reconstitute the negative frequencies, this
module effects a 4-in-1 real biplex FFT that can then be substituted for the
equivalent number of biplex cores in a high-bandwidth FFT.  Thus, our real FFT
module has the same bandwidth flexibility as our standard complex FFT.

Dynamic range inside fixed-point FFTs requires careful consideration.  Tones
are folded into half as many samples through each FFT stage, causing magnitudes
to grow by a factor of 2 for narrow-band signals, and $\sqrt{2}$ for random 
noise.  To
avoid overflow and spectrum corruption, our cores contain optional downshifts
at each stage.  In an interference-heavy environment, one must balance loss of
SNR from downshifting signal levels against loss of integration time due to
overflows.  A good practice is to place time-domain input into the
most-significant bits of the FFT and downshift as often as possible to
avoid overflow and minimize rounding error in each butterfly stage.  However,
it is also best to avoid using the top 2 bits on input since the first 
2 butterfly
stages can be implemented using negation instead of complex multiplies, but the
asymmetric range of 2's complement arithmetic can allow this negation to
overflow.

\subsection{A Cross-Multiplication/Accumulation (X) Engine}
\label{sec:x_engine_arch}

\begin{figure}
\begin{center}
\includegraphics[scale=.45]{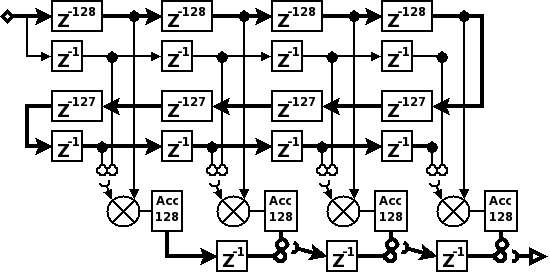}
\caption{%
This X engine schematic illustrates the pipelined flow of data
that allows it to be split across multiple FPGAs and boards.
With continuous data input, all multipliers (with the possible exception of
the final stage for even values of $N_{ant}$) are used with 100\% efficiency.
\label{fig:x_engine_schem}}
\end{center}
\end{figure}

\placefigure{fig:x_engine_schem}

Our FX correlator architecture employs
X engines to compute all antenna cross-multiples within a frequency
channel, and multiple frequencies are multiplexed into the core as dictated by
processor bandwidth; the complex visibility $V_{ij}$ (Eq. \ref{eq:vis})
is the average of the product of complex voltage samples from antenna $i$ and
antenna $j$ with the convention that the voltage $j>i$ is conjugated prior to
forming product.
In collaboration with Lynn Urry of UC Berkeley's Radio
Astronomy Lab we have implemented a parametrized module (Fig.
\ref{fig:x_engine_schem}) for computing and accumulating all visibilities for a
specified number of antennas.  An X engine operates by receiving $N_{ant}$ data
blocks in series, each containing $T_{acc}$ data samples from one frequency
channel of one antenna.  The first samples of all blocks are
cross-multiplied, and the $N_{ant}(N_{ant}+1)/2$ results are added to the
results from the second samples, and so on, until all $T_{acc}$ samples have
been exhausted.  Accumulation prevents the data rate out of a
cross-multiplier from exceeding the input data rate.  An X engine is divided
into stages, each responsible for pairing two different data blocks
together: the zeroth stage pairs adjacent blocks, the first stage pairs blocks
separated by one, and so on.  As the final accumulated results become available,
they are loaded onto a shift register and output from the X engine.

However, as a new window of $N_{ant}\times T_{acc}$ samples arrives, some
stages, behaving as described above, would compute invalid results using
data from two different windows.  To avoid this, each stage switches between
cross-multiplying separations of $S$ to separations of $N_{ant}-S$, which
happen to be valid precisely when separations of $S$ would be invalid.  As a
result, there need be only $floor({N_{ant}/2}+1)$ stages in an X engine.  Every
$T_{acc}$ samples, each stage outputs a valid result, yielding $N_{ant}\times
floor({N_{ant}/2}+1)$ total accumulations; for even values of $N_{ant}$,
$N_{ant}/2$ of the results from the last stage are redundant.
All other multiplier/accumulators are 100\% utilized.  Each stage
also computes all polarization cross-multiples (Eq. \ref{eq:pol})
using parallel multipliers.

When one X engine no longer fits on a single FPGA, it may be divided across
chips at any stage boundary at the cost of a moderate amount of bidirectional
interconnect.  The output shift register need not be carried between chips;
each FPGA can accumulate and store the results computed locally.  In order for
the output shift register's $floor({N_{ant}/2}+1)$ stages to clear before the
next accumulation is ready, an X engine requires a minimum integration length
of: $T_{acc}>floor({N_{ant}/2}+1)$.  In current hardware, a practical upper
limit on $T_{acc}$ is set by the 2$\times$4 Mbit of SRAM storage available on
the IBOB.  For 2048 channels with 4-bit samples, and double buffering for 2
antennas, 2 polarizations, this limit is $T_{acc}\le 128$.  Longer integration
requires an accumulator capable of buffering an entire vector of visibility
data, and typically occurs in off-chip DRAM.  The maximum theoretical
accumulation length in correlator is determined by the fringe rate of sources
moving across the sky, and is a function of observing frequency, maximum
antenna separation, and (for correlators with internal fringe rotation)
field-of-view across the primary beam.

Cross-multiplication comes to dominate the total correlator processing budget
for large numbers of antennas.  As a result, care must be taken both to reduce
the footprint of a complex multiplier/accumulator and to make full and
efficient use of the resources on an FPGA processor.  The number of bits used
to carry a signal should be minimized while retaining sufficient dynamic range
to distinguish signal from noise.  We have chosen to focus on 4-bit multipliers
in current applications, and the subjects of dynamic equalization and Van Vleck
correction generalized to 4 bits are explored in Section
\ref{sec:characterization} for optimizing signal-to-noise ratios (SNR) in our
correlators.  To make full use of FPGA resources, we construct
4-bit complex multipliers using distributed logic, dedicated multiplier cores, 
and look-up tables implemented in Block RAMs.  

It is possible to perform the bulk of an $N$-bit complex multiply in an $M$-bit
multiplier core by sign-extending numbers to $2N$ bits and combining them into
two $M$-bit, unsigned numbers.  Multiplying $(a+bi)(c+di)$, these
representations are $(2^{M-2N}a_s+b_s)$ and $(2^{M-2N}c_s+d_s)$, where
$n_s=2^{2N}+n$.  The bits corresponding to $ac, ad+bc, bd$ may be selected from
the product, provided that the
sign-extension to $2N$ bits shifts $a+d$ beyond the bits occupied by $ad$.
This yields the constraint: 
\begin{equation} 6N-1 < M \end{equation} 
The 18-bit multipliers in current Xilinx 
FPGAs can efficiently perform 3-bit complex
multiplies, but fall short of 4 bits.

\section{System Integration}
\label{sec:integration}

\subsection{F Engine Synchronization}
\label{sec:F_synch}

\begin{figure}
\begin{center}
\includegraphics[scale=.45]{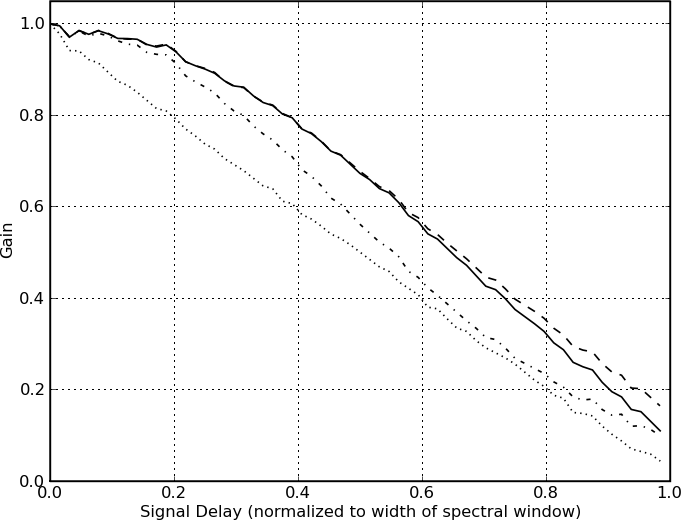}
\caption{%
Cross-correlation of noise decreases as a function of signal delay between 
antenna inputs.  PFBs operate on a wider window of data compared to DFTs, and 
use non-flat sample weightings, yielding a
different correlation response versus signal delay compared to the standard
result presented in Thompson et al. (2001) \cite{thompson_et_al2001}.  Graphed
are the responses of PFBs with 8 taps (solid), 4 taps (dashed), 2 taps (dot
dashed), and the response of a DFT (dotted).
\label{fig:corr_vs_dly}}
\end{center}
\end{figure}

\placefigure{fig:corr_vs_dly}

Though we have touted GALS design principles for X engine processing,
digitization and spectral processing within F engines must be synchronized to a
time interval much smaller than a spectral window to avoid severe degradation
of correlation response (Fig. \ref{fig:corr_vs_dly}).  This attenuation effect,
resulting from the changing degree of overlap of correlated signals within a
spectral window, can be caused by systematic signal delay between antennas, as
well as by source-dependent geometric delay; FX correlators with insufficient
channel resolution experience a narrowing of the field of view related to
channel bandwidth.  This effect has been well explored for FX correlators
employing DFTs (see Chapter 8 of \citet{thompson_et_al2001}), but Polyphase
Filter Banks show a different response owing to a weighting function that
extends well beyond the number of samples used in a DFT. 
Given a standard form for PFB sample weighting of
${\rm sinc}\left(\frac{\pi t}{N\tau_s}\right)
W\left(\frac{t}{2TN\tau_s}\right)$, 
where $N$ is the number of output channels,
$T$ is the number of PFB taps, $\tau_s$ is the delay between time-domain
samples, and $W$ is an arbitrary windowing function that tapers to 0 at
$\pm1$, the gain versus delay $G(\tau)$ of a PFB-based FX correlator is
given by:
\begin{displaymath}
G(\tau)=\int_{-\infty}^{\infty}{
\left[{\rm sinc}\left(\frac{\pi t}{N\tau_s}\right)
W\left(\frac{t}{2TN\tau_s}\right)\right] \times
\left[{\rm sinc}\left(\frac{\pi (t-\tau)}{N\tau_s}\right)
W\left(\frac{t-\tau}{2TN\tau_s}\right)\right]\ dt
}
\end{displaymath}

For the purpose of F Engine synchronization, we
rely on a one-pulse-per-second (1PPS) signal with a fast edge-rate provided
synchronously to a bank of F processors running off identical system clocks.
This signal is sampled by the system clock on each processor, and provided
alongside ADC data.  A slower, asynchronous ``arm'' signal is sent from
a central node to each F engine at the half second phase 
to indicate that the next 1PPS signal should be
used to generate the reset event that synchronizes spectral windows and packet
counters.  This ensures that samples from different antennas entering X engines
together were acquired within one or two system clocks of one another.  The
degree of synchronization is determined by the difference in path lengths of
1PPS and the system clock from their generators to each F engine.  This path
length can be determined from celestial source observations
using self-calibration, and barring temperature
effects, will be constant for a correlator configuration following power-up.

\subsection{Asynchronous, Packetized ``Corner Turner''}
\label{sec:packetization}

The choice of the accumulation length $T_{acc}$ in X engines 
determines the natural size of UDP packets in our
packet-switched correlator architecture.  For current CASPER hardware where
channel-ordering occurs in IBOB SRAM, $T_{acc}$ is constrained by the available
memory to an upper limit of 128 samples for 2048-channel dual-polarization, 
4-bit,
complex data, yielding a packet payload of 256 bytes.  A header containing
2 bytes of antenna index and 6 bytes of frequency/time index is added to each
packet to enable packet unscrambling on the receive side.  The frequency/time
index (hereafter referred to as the master counter, or MCNT) is a counter that
is incremented every packet transmission.  The lower bits count frequencies
within a spectrum, and the rest count time.  Combined with the antenna
index, MCNT completely determines the time, frequency, source, and destination
of each packet; MCNT maps uniquely to a destination IP address.

\begin{figure}
\begin{center}
\includegraphics[scale=.5]{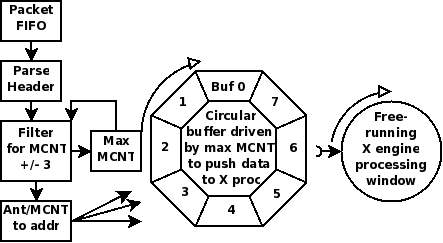}
\caption{Before transmission, each F engine packet is 
tagged with an antenna number and master counter (MCNT) encoding
time and frequency.  Received packets are filtered to 
the narrow range of MCNTs, and maximum MCNT slides smoothly up as packets 
are received.  A free-running X engine 
processes available windows when it is ready.  This architecture
allows data to be processed at a lower data rate than the FPGA clock rate 
without requiring every element in the pipeline to have a enable signal.
\label{fig:packet_rx}}
\end{center}
\end{figure}

\placefigure{fig:packet_rx}

Packet reception (Fig. \ref{fig:packet_rx}) is complicated by the realities of
packet scrambling, loss, and interference.  A circular buffer holding $N_{win}$
windows worth of X engine data stores packet data as they arrive.  The lower
bits of MCNT act as an address for placing payloads into the the correct
window, and the antenna index addresses the position within that window.  When
data arrives $N_{win}/2$ windows ahead of a buffered window, that window is
flagged for readout, and is processed contiguously on the next window boundary
of the free-running X engine.  Using packet arrival to determine when a window
is processed allows a data-rate dependent time interval for all packets to
arrive, but pushes data through the buffer in the event of packet loss.  On
readout, the buffer is zeroed to ensure that packet loss results in loss of
signal, rather than the introduction of noise.  F engines can be intentionally
disconnected from transmission without compromising the correlation of
those remaining.

Packet interference occurs when a well-formed packet contains an invalid MCNT
as a result of switch latency, unsynchronized F engines, or system
misconfiguration.  Such packets must be prevented from entering the receive
buffer, since they can lead to data corruption; one would prefer that a
misconfigured F engine antenna result in data loss for that antenna, rather
than data loss for the entire system.  To ensure this behavior, incoming
packets face a sliding filter based on currently active MCNTs.  Packets are
only accepted if their MCNT falls within the range of what can currently be
held in the circular buffer.  As higher MCNTs are received and accepted, old
windows are flagged for read out, freeing up buffer space for still
higher MCNTs.  This system forces MCNTs to advance by small increments and
prevents the large discontinuities indicative of packet
interference.  In the eventuality that a receive buffer accidentally locks onto
an invalid MCNT from the outset, a time-out clause causes the currently active
MCNT to be abandoned for a new one if no new data is accepted into the receive
buffer.

A final complication comes when implementing a bidirectional 10GbE transmission
architecture such as the one outlined in Figure \ref{fig:ex_app1}.
Commercial switches do not support
self-addressed packet transmission; they assume that the transmitter
(usually a CPU) intercepts these packets and transfers them to the receive
buffer.  On FPGAs, this requires an extra buffer for holding ``loopback'', and
a multiplexer for inserting these packets into the processing stream.  A simple
method for this insertion would be to always insert loopback packets, if
available, and otherwise to insert packets from the 10GbE
interface.  However, there is a maximum interval over which packets with
identical MCNTs can be scrambled before the receive system rejects
packets for being outside of its buffer.  This simple method has the
undesirable effect of including switch latency in the time interval over which
packets are scrambled, causing unnecessary packet loss.  Our solution is to
pull loopback packets only after packets with the same MCNT 
arrive through the switch.

\subsection{Monitor, Control, and Data Acquisition}
\label{sec:data_aq}

The toolflow we have developed for CASPER hardware provides convenient
abstractions for interfacing to hardware components such as ADCs, DRAM, and 10
GbE transceivers, and allows specified registers and BRAMs to be automatically
connected to CPU-accessible buses.  On top of this framework, we run BORPH--an
extension of the Linux operating system that provides kernel support for FPGA
resources \citep{so_broderson2006,so2007}.  This system allows FPGA
configurations to be run in the same fashion as software processes, and creates
a virtual file system representing the memories and registers defined on the
FPGA.  Every design compiled with this toolflow comes equipped with this
real-time interface for low- to moderate-bandwidth data I/O.  By emulating
standard file-I/O interfaces, BORPH allows programmers to use standard
languages for writing control software.  The majority of the monitor, control,
and data acquisition routines in our correlators are written in C
and Python.  For 8-16 antenna correlators, the bandwidth through BORPH on a
BEE2 board is sufficient to support the output of visibility data with 5-10s
integrations.

For correlators with more antennas or shorter integration times, the bandwidth
of the CPU/FPGA interface is incapable of maintaining the full correlator
output.  This limitation is being overcome by transmitting the final correlator
output using a small amount of the extra bandwidth on the 10GbE ports already
attached to each X engine.  After accumulation in DRAM, correlator output is
multiplexed onto the 10GbE interface and transmitted to one or more Data
Acquisition (DA) systems attached to the central 10GbE switch.  These systems
collect and store the final correlator output.  With a capable DA system, the
added bandwidth through this output pathway can be used to attain millisecond
integration times, opening up opportunities for exploring transient events and
increasing time resolution for removing interference-dominated data. 

The capabilities of correlators made possible by our research are placing
new challenges on DA systems \citep{wright2005}.  There is a severe (factor of
100) mismatch between the data rates in the on-line correlator hardware and
those supported by the off-line processing.  Members of our team are currently
pursuing research on how this can be resolved both for correlators and for
generic signal processing systems using commercially available compute
clusters.  For correlators, our group is currently exploring how to implement
calibration and imaging in real-time to reduce the burden of
expert data reduction on the end user, and to make best use of both telescope
and human resources.

\section{Characterization}
\label{sec:characterization}

\subsection{ADC Crosstalk}
\label{sec:crosstalk}

\begin{figure}
\begin{center}
\includegraphics[scale=.35]{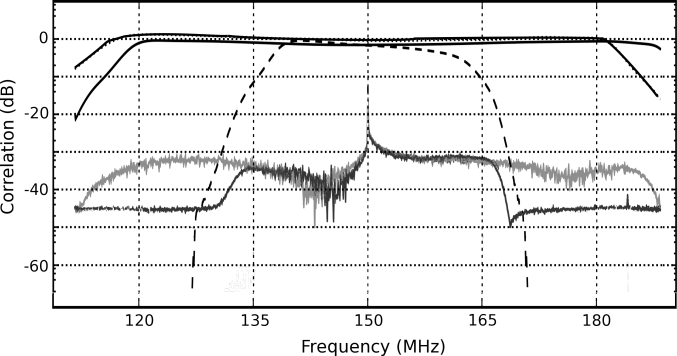}
\caption{%
Uncorrelated noise sources with similar bandpass shapes were
input to two channels of one ADC board (solid black) and a third noise source 
with a narrower passband was input to to a second ADC board 
(dashed black) in the ``Pocket Correlator'' system.
Crosstalk levels between signal inputs on the same ADC board (light gray) and 
between ADC boards sharing an IBOB (dark gray) peak at $-28$ dB.
\label{fig:crosstalk}}
\end{center}
\end{figure}

\placefigure{fig:crosstalk}

Crosstalk is an undesirable but prevalent characteristic of analog systems
wherein a signal is coupled at a low level into other pathways.  This can pose
a major threat to sensitivity in systems that integrate noise-dominated data to
reveal low-level correlation.  For CASPER hardware, we have examined crosstalk
levels between signal inputs sharing an ADC chip, and between different ADC
boards on the same IBOB.  Figure \ref{fig:crosstalk} illustrates a one-hour
integration of uncorrelated noise of various bandwidths input to the ``Pocket
Correlator'' system (see Section \ref{sec:deployments}).  Between inputs 
of the same ADC board, a coupling coefficient of $\sim0.0016$ indicates
crosstalk at approximately $-28$ dB.  This coupling is a factor of $5$ higher
than the $-35$ dB isolation advertised by the Atmel ADC chip, and is most
likely the result of board geometry and shared power supplies.  Crosstalk
between inputs on different ADCs also peaks at the $-28$ dB level, but shows
more frequency-dependent structure.

\begin{figure}
\begin{center}
\includegraphics[scale=.5]{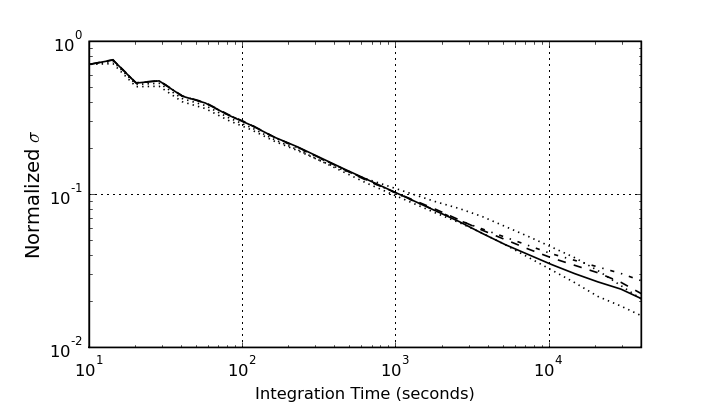}
\caption{%
Measurements of the standard deviation versus integration time of the 
correlation between independent noise sources into the same ADC board show 
that crosstalk exhibits 
stability over a period of 1 day for all frequency channels
Although phase switching
may still be desireable, this stability allows
crosstalk to be calibrated and removed after correlation.
\label{fig:crosstalk_stability}}
\end{center}
\end{figure}

\placefigure{fig:crosstalk_stability}

Crosstalk may be characterized and removed, provided that its timescale for
variation is much longer than the calibration interval.  Figure
\ref{fig:crosstalk_stability} demonstrates that for integration intervals
ranging from 7.15 seconds to approximately 1 day (the limit of our testing),
crosstalk amplitudes and phases vary around stable values in a
lab test that, when
subtracted, yield noise that integrates down with time.  Even
though crosstalk is encountered at the $-28$ dB level, its stability allows
suppression to at least $-62$ dB.  This stability has allowed crosstalk
to be removed post-correlation, and we have until recently deferred
adding phase switching.  Developments along this line are proceeding by
introducing an invertible mixer (controlled via a Walsh counter on an IBOB)
early in the analog signal path, and removing this inversion after
digitization.  Phase switching must be coupled with data blanking near 
boundaries when the
inversion state is uncertain.  Blanking will be most easily implemented by
intentionally dropping packets of data from F engine transmission, and by
providing a count of results accumulated in each integration for normalization
purposes.

\subsection{XAUI Fidelity and Switch Throughput}
\label{sec:10gbe_sw}

CASPER boards are currently configured to transmit XAUI protocol over CX4 ports
as a point-to-point communication protocol and as the physical layer of 10GbE
transmission.  Because the Virtex-II FPGAs used in current CASPER hardware do
not fully support XAUI transmission standards \cite{xilinx_ug024,xilinx_ds083}, 
current devices can have
sub-optimal performance for certain cable lengths.  We expect the new ROACH
board, which employs Virtex-5 FPGAs, to have better
performance in this regard.  For cable lengths supported in current hardware,
we tested XAUI transmission fidelity using matched Linear Feedback Shift
Registers (LFSRs) on transmit and receive.  Error detection was verified using
programmable bit-flips following transmitting LFSRs.  Over a period of 16
hours, 573 Tb of data were transmitted and received on each of 8 XAUI
links.  During this time, no errors were detected, resulting in an estimated
bit-error rate of $2.2\cdot 10^{-16}$ Hz.  We also tested the capability of two
Fujitsu switches (the XG700 and the XG2000) for performing the full
cross-connect packet switching required in our FX correlator architecture.  By
tuning the sample rate inside F engines of an 8-antenna (4-IBOB) packetized
correlator, we controlled the transmission rate per switch port over a range of
5.96 to 8.94 Gb/s.  In 10-minute tests, packet loss was zero for both
switches in all but the 8.94 Gb/s case.  Packet loss in this final case was
traced to intermittent XAUI failure as a result of imperfect compliance with
the XAUI standard, as described previously. Overheating of FPGA chips in the
field has also been reported as a source of intermittent operation.

\subsection{Equalization and 4-Bit Requantization}
\label{sec:equalization}

\begin{figure}
\begin{center}
\includegraphics[scale=.52]{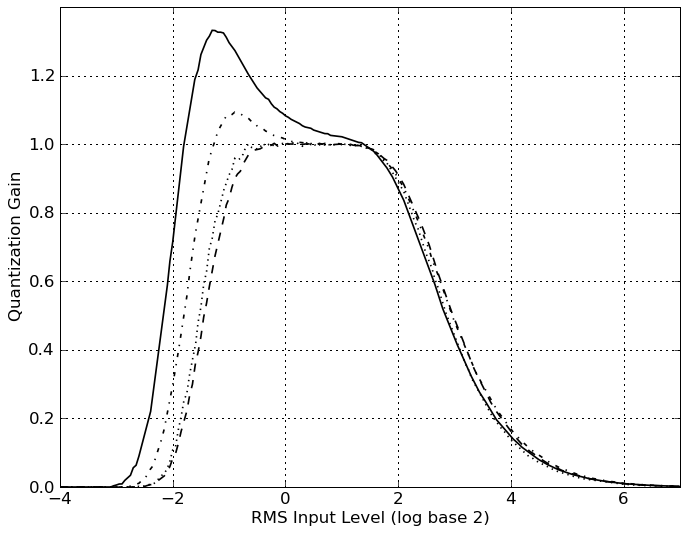}
\caption{%
Illustrated above is the relative gain through a 4-bit, 15-level quantizer as a 
function of input signal level (log base 2).  Plotted are gain curves for 
the cross-correlation of two
gaussian noise sources with correlation levels of 100\% (solid),
80\% (dot-dashed), 40\% (dotted), and 20\% (dashed).  
\label{fig:4_bit_quant}}
\end{center}
\end{figure}

\placefigure{fig:4_bit_quant}

Correlator processing resources can be reduced by limiting the bit width of
frequency-domain antenna data before cross-multiplication.  However, digital
quantization requires careful setting of signal levels for optimum
SNR and subsequent calibration to a linear power scale 
\citep{thompson_et_al2001,jenet_anderson1998}.  Correlators using 4 bits 
represent
an improvement over their 1 and 2 bit predecessors, but there are still
quantization issues to consider.  The total power of a 4-bit quantizer has a
non-linear response with respect to input level as shown in Figure
\ref{fig:4_bit_quant}.  In currently deployed correlators, we perform
equalization (per channel scaling) to control the RMS channel values before
requantizing from 18 bits to 4 bits.  This operation saturates RFI and flattens
the passband to reduce dynamic range and to hold the passband in
the linear regime of the 4-bit quantization power curve.  Equalization is
implemented as a scalar multiplication on the output of each PFB using 18-bit
coefficients from a dynamically updateable memory.  These coefficients allow
for automatic gain control to maintain quantization fidelity through changing
system temperatures.

\section{Deployments and Results}
\label{sec:deployments}

\subsection{A Pocket Correlator}
\label{sec:pocket_corr}

\begin{figure}
\begin{center}
\includegraphics[scale=.5]{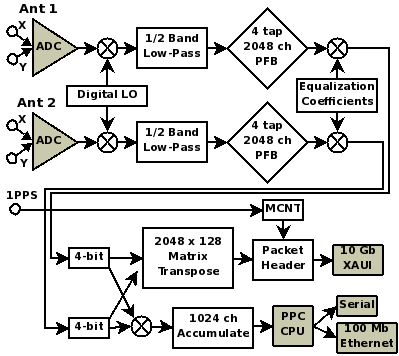}
\caption{%
This IBOB design serves a dual purpose as a stand-alone ``Pocket Correlator''
and an F processor in a 16 antenna packetized correlator deployment.  Note the
parallel output pathways for each function.
\label{fig:f_engine}}
\end{center}
\end{figure}

\placefigure{fig:f_engine}

The ``Pocket Correlator'' (Fig. \ref{fig:f_engine}) is a single IBOB system
that includes F and X engines on a single board for correlating and
accumulating 4 input signals.  Each input is sampled at 4 times the FPGA clock
rate (which runs up to 250 MHz), and a down-converter extracts half of the
digitized band.  This subband is decomposed into 2048 channels by an 8-tap PFB,
equalized, and requantized to 4 bits.  With all input signals on one chip, X
processing can be implemented directly as multipliers and vector accumulators,
rather than as X engines.  Limited buffer space on the IBOB permits only 1024
channels (selectable from within the 2048) to be accumulated.  Output occurs
either via serial connection (with a minimum integration time of 5
seconds) or via 100-Mbit UDP transmission (with a minimum integration time in
the millisecond range).  This system can act as a 2-antenna, full Stokes
correlator, or as a 4-antenna single polarization correlator.

\begin{figure}
\begin{center}
\includegraphics[scale=.25]{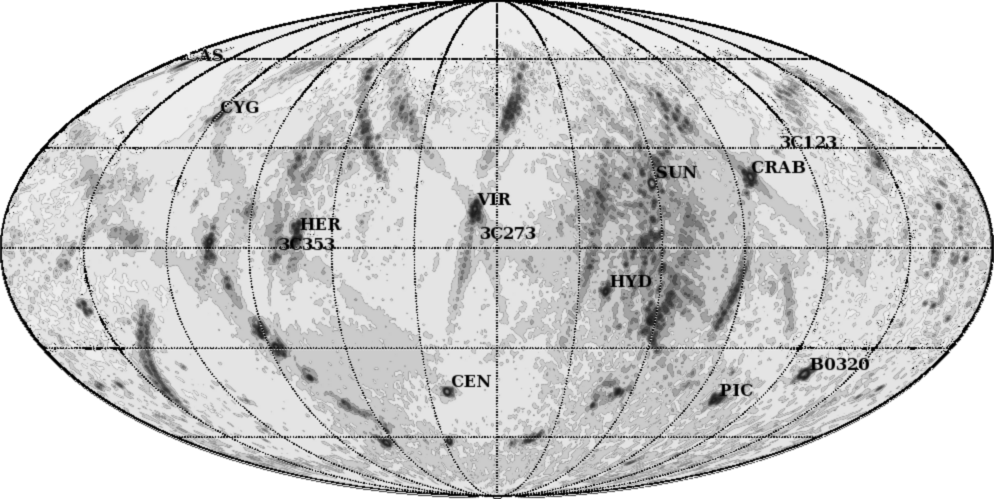}
\caption{%
This all-sky image, made using a 75-MHz band centered at 150 MHz with the
``Pocket Correlator'' as part of the PAPER experiment in Western 
Australia, achieves an impressive 10,000:1 signal-to-noise ratio using 
1 day of data. 
\label{fig:skymap}}
\end{center}
\end{figure}

\placefigure{fig:skymap}

The Pocket Correlator is valuable as a simple, stand-alone instrument, and for
board verification in larger packetized systems.  It is being applied as a
stand-alone instrument in PAPER, the ATA, and the UNC PARI observatory. A
4-antenna, single polarization deployment of the PAPER experiment in Western
Australia in 2007 used the Pocket Correlator to collect the data used to
produce a 150 MHz all-sky map illustrated in Figure \ref{fig:skymap}.  In
addition to demonstrating the feasibility of post-correlation crosstalk
removal, this map (specifically, the imperfectly removed sidelobes of sources)
illustrates a problem that will require real-time imaging to solve for large
numbers of antennas.

\subsection{An 8-Antenna, 2-Stokes, Synchronous Correlator}
\label{sec:8_ant_corr}

This first generation multi-board correlator demonstrated the functionality
of signal processing algorithms and CASPER hardware, but preempted the
current packetized architecture--it operated synchronously.  This version of
the correlator was most heavily limited by X engine resources, all of which
were implemented on a single FPGA to simplify interconnection. The
total number of complex multipliers in the X engines of an $N_{ant}$ antenna
array is: $N_{cmac} = floor({N_{ant}/2}+1)\times N_{ant}\times N_{pol}$; the
limited number of multipliers on a BEE2 FPGA only allowed for supporting half
the polarization cross-multiples.  This system was an
important demonstration of the basic capabilities of our hardware and software,
and provided a starting point for evolving a more sophisticated system.  
Deployments of this
system at the NRAO site in Green Bank as part of the PAPER
experiment, and briefly
at the Hat Creek Radio Observatory for the Allen Telescope Array,
are being supersede by the packetized correlator presented in the next
section.

\subsection{A 16-Antenna, Full-Stokes, Packetized Correlator}
\label{sec:packet_deploy}

This packetized FX correlator is a realization of the architecture outlined in
Figure \ref{fig:ex_app1}, with F processing for 2 antennas implemented on each
IBOB, and matching X processors implemented on each corner FPGA of two BEE2s.
Each F processor is identical to a Pocket Correlator (Fig. \ref{fig:f_engine}),
but branches data from the equalization module to a matrix transposer in IBOB
SRAM to form frequency-based packets.  Packet data for each antenna are
multiplexed through a point-to-point XAUI connection to a BEE2-based X
processor, and then relayed in 10GbE format to the switch.  The number of
channels in this system is limited to 2048 by memory in IBOB SRAM for
transposing the 128 spectra needed to meet bandwidth restrictions between X
engines and DRAM-based vector accumulators.

\begin{figure}
\begin{center}
\includegraphics[scale=.5]{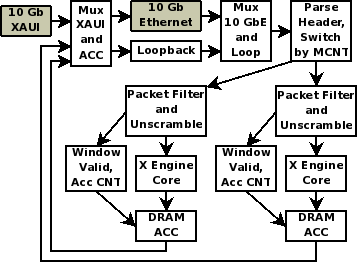}
\caption{%
A BEE2-based X processor in a packetized correlator transmits data 
from an F engine
over 10GbE and stores self-addressed packets in a ``loopback'' buffer.
These streams are merged on the receive side, and packets are
distributed to two X engines.  Accumulation occurs
in DRAM buffers, and the results are packetized and output
over the same 10GbE link.  A data aquisition system connects to the
same switch as the X engines.
\label{fig:x_processor}}
\end{center}
\end{figure}

\placefigure{fig:x_processor}

The X processor in this packetized correlator implements the transmit and
receive architecture illustrated in Figure \ref{fig:x_processor} 
for two X engines sharing the same 10GbE link.
Each X engine's data processing rate is
determined by packets arriving in its own receive buffer, and results are
accumulated in separate DRAM DIMMs.  The accumulated output of each X processor
is read out of DRAM at a low bandwidth and transmitted via 10GbE packets to
a CPU-based server where
all visibility data is collected and
written to disk in MIRIAD format
\citep{sault_et_al1995} using interfaces from the Astronomical Interferometry
in PYthon (AIPY) package\footnote{http://pypi.python.org/pypi/aipy}.

The clocks for the BEE2 FPGAs are asynchronous 200-MHz oscillators, and IBOBs
run synchronously at any rate lower than this.  Packet transmission is
statically addressed so that all each X engine processes every 16th channel.
We use 8 ports of a Fujitsu XG700 switch to route data.  This system is is
scalable to 32 antennas before two X engines no longer fit on a single FPGA.
For larger systems, the number of BEE2s will scale as the square of the number
of antennas, and the number of IBOBs will scale linearly.  A 32-antenna,
200-MHz correlator on 16 IBOBs and 4 BEE2s is now working in the lab, and a
16-antenna version using 8 IBOBs and 2 BEE2s has been deployed to the NRAO site
in Green Bank with the PAPER experiment.  

\section{Conclusion}
\label{sec:conclusion}

By decreasing the time and engineering costs of building and upgrading
correlators, we aim to reduce the total cost of correlators for a wide range of
scales.  Small- and medium-scale correlators with total cost dominated by
development clearly stand to benefit from our research.  It is less clear if
the cost of large-scale correlators can be reduced by the general-purpose
hardware used in our architecture.  Though minimization of replication cost
favors the development of specialized parts, there are two factors
that can make a generic, modular solution cost less.

The first factor to consider is time to deployment.  Even if the monetary cost
of development is negligible in the budget of a large correlator, the cost of
development time can be significant.  If a custom solution takes several years
to go from design to implementation, the hardware that is deployed will be out
of date.  Moore's Law suggests that when a custom solution taking 3 years to
develop is deployed, there will exist processors 4 times more powerful, or 4
times less expensive for the equivalent system.  The cost of a generic, modular
system has to be tempered by the expected savings of committing to hardware
closer to the ultimate deployment date.

The second factor is the cost of upgrade.  Many facilities (including the ATA)
are beginning to appreciate the advantages of designing arrays with wider
bandwidths and larger numbers of antennas than can be handled by current
technology.  Correlators may then be implemented inexpensively on scales
suited to current processors, and upgraded as more powerful processors
become available.  Modular solutions facilitate this methodology.


\acknowledgments

This and other CASPER research are supported by the National Science Foundation
Grant No. 0619596 for Low Cost, Rapid Development Instrumentation for Radio
Telescopes.  We would like to acknowledge the students, faculty and sponsors of
the Berkeley Wireless Research Center, and the National Science Foundation
Infrastructure Grant No.  0403427.  Correlator development for the PAPER
project is supported by NSF grant AST-0505354, and for the ATA project by NSF
grant AST-0321309 as well as the Paul G. Allen Foundation.  Chips and software
were generously provided by Xilinx, Inc.  JM and PM gratefully acknowledge
financial support from the MeerKAT project and South Africa's National Research
Foundation.

\appendix
Glossary of Technical Terms
\begin{itemize}
\item ADC - Analog to Digital Converter
\item ASIC - Application-Specific Integrated Circuit processor
\item BEE2 - Berkeley Emulation Engine, rev. 2
\item BORPH - Berkeley Operating system for Re-Programmable Hardware
\item BRAM - Block RAM: Random Access Memory inside an FPGA
\item CX4 - 10GbE-compatible industry standard connector
\item CPU - Central Processing Unit
\item DDR2 - Double-Data-Rate 2 type of off-FPGA Synchronous DRAM 
\item DIMM - Dual Inline Memory Module
\item DFT - Discrete Fourier Transform
\item DRAM - Dynamic Random Access Memory
\item FFT - Fast Fourier Transform algorithm
\item FIR - Finite Impulse Response digital filter
\item FPGA - Field Programmable Gate Array processor
\item FX - Correlator architecture implemented as frequency channelization, then cross-multiplication
\item GALS - Globally Asynchronous, Locally Synchronous system architecture
\item GB - GigaByte
\item IBOB - Internet Break-Out Board
\item LFSR - Linear Feedback Shift Register
\item LO - Local Oscillator
\item MCNT - Master Counter
\item PFB - Polyphase Filter Bank
\item PowerPC - a specific CPU architecture
\item QDR - Quad-Data-Rate type of off-FPGA SRAM
\item ROACH - Reconfigurable, Open Architecture for Computing Hardware
\item SNR - Signal-to-Noise Ratio
\item SRAM - Static Random Access Memory
\item UDP - User Datagram Protocol Ethernet packetization
\item XAUI - X (ten) Attachment Unit Interface point-to-point transmission protocol
\item XF - Correlator architecture implemented as cross-multiplication, then frequency channelization
\item 1PPS - 1 Pulse Per Second clock signal
\item 10GbE - 10 Gigabit per second Ethernet communication standard
\end{itemize}

\clearpage

\begin{table}[t]
\label{tab:hardware_price}
\begin{center}
\title{Price and Power Consumption of CASPER Hardware}
\begin{tabular}{lrrrrr}
\hline\hline
\vspace{3pt}
Board & Board & Cost with & Gops    & Power \\
      & Cost  & FPGAs     & per Sec & (W)\\
\hline
IBOB& \$400 & \$2700 & 70 & 30 \\
BEE2& \$5000 & \$23500 & 500 & 150 \\
ROACH$^*$& \$1000 & \$3200 & 400 & 50 \\
ADC (1Gs/s$\times2$)& \$200 & \$200 & N/A & 2 \\
ADC (3Gs/s)\tablenotemark{*}& \$1000 & \$1000 & N/A & 5 \\
\hline\hline
\vspace{-5pt}
\end{tabular}
\\
\vspace{-10pt}
\tablenotetext{*}{Estimated from prototype versions.}
\end{center}
\end{table}


\end{document}